\documentclass[aps,prl,twocolumn,superscriptaddress,longbibliography]{revtex4-2}
\usepackage[colorlinks=true, citecolor=blue, urlcolor=blue, linkcolor=red]{hyperref}
\usepackage[english]{babel}
\usepackage{amssymb,amsmath,orcidlink,physics}
\renewcommand{\section}[1]{{\par\it #1.---}\ignorespaces}

\begin{document}
\title{Topological phases in non-Hermitian nonlinear-eigenvalue systems}
\author{Yu-Peng Ma\orcidlink{0009-0006-9911-9424}}
\affiliation{Key Laboratory of Quantum Theory and Applications of Ministry of Education, Lanzhou Center for Theoretical Physics, Gansu Provincial Research Center for Basic Disciplines of Quantum Physics, Key Laboratory of Theoretical Physics of Gansu Province, Lanzhou University, Lanzhou 730000, China}
\author{Ming-Jian Gao\orcidlink{0000-0002-6128-8381}}
\affiliation{Key Laboratory of Quantum Theory and Applications of Ministry of Education, Lanzhou Center for Theoretical Physics, Gansu Provincial Research Center for Basic Disciplines of Quantum Physics, Key Laboratory of Theoretical Physics of Gansu Province, Lanzhou University, Lanzhou 730000, China}
\author{Jun-Hong An\orcidlink{0000-0002-3475-0729}}\email{anjhong@lzu.edu.cn}
\affiliation{Key Laboratory of Quantum Theory and Applications of Ministry of Education, Lanzhou Center for Theoretical Physics, Gansu Provincial Research Center for Basic Disciplines of Quantum Physics, Key Laboratory of Theoretical Physics of Gansu Province, Lanzhou University, Lanzhou 730000, China}

\begin{abstract}
The discovery of topological phases has ushered in a new era of condensed matter physics and revealed a variety of natural and artificial materials. They obey the bulk-boundary correspondence (BBC), which guarantees the emergence of boundary states with nonzero topological invariants in the bulk. Widespread attention has been paid to extending topological phases to nonlinear and non-Hermitian systems. However, the BBC and topological invariants of non-Hermitian nonlinear systems  remain largely unexplored. Here, we establish a complete BBC and topological characterization of the topological phases in a class of non-Hermitian nonlinear-eigenvalue systems by introducing an auxiliary system. We restore the BBC broken by non-Hermiticity via employing the generalized Brillouin zone on the auxiliary system. Remarkably, we discover that the interplay between non-Hermiticity and nonlinearity creates an exotic complex-band topological phase that coexists with the real-band topological phase. Our results enrich the family of nonlinear topological phases and lay a foundation for exploring novel topological physics in metamaterial systems.
\end{abstract}
\maketitle
	
\section{Introduction}
The rapid development of topological insulators and superconductors has triggered a revolution in condensed matter physics and unlocked diverse advanced functionalities in natural and artificial materials \cite{RevModPhys.82.3045,RevModPhys.83.1057,RevModPhys.88.021004,RevModPhys.88.035005,RevModPhys.89.040502,PhysRevX.7.041069}. Going beyond Landau's phase transition paradigm, they are not caused by the breaking of symmetries \cite{PhysRevLett.95.226801,PhysRevLett.95.146802}. Their common feature is the presence of boundary states in the bulk gap, which are protected by symmetries and robust to disorder. What lies at the heart of them is the bulk-boundary correspondence (BBC). It indicates that these boundary states are intrinsically guaranteed by the topology of the bulk bands of systems \cite{PhysRevB.78.195125,PhysRevB.76.045302,PhysRevLett.132.136401,Zhu2025,Li2024,Hossain2024,PhysRevX.14.041048,PhysRevB.102.115135}. Depending on different symmetries satisfied by the bulk bands, topological phases are classified into various categories \cite{RevModPhys.88.035005,PhysRevB.78.195125,PhysRevResearch.6.033192}. This classification rule has not only guided the discovery of rich topological phases of matter in different electronic materials \cite{Zhang2025,doi:10.1126/science.adk1270,doi:10.1126/science.adj3742,Bouhon2020}, but also the simulation of these phases on multiple platforms, including photonics \cite{Lu2014,Huang2024,doi:10.1126/science.adr5234,33mm-mx88}, circuit \cite{PhysRevX.5.021031,Lenggenhager2022,Wang2020}, ultracold atom \cite{Jotzu2014,PhysRevLett.131.263001,Braun2024}, and acoustic \cite{Kane2014,Xue2019,Cheng2025,PhysRevLett.132.216602} systems.

There has been widespread interest in the topological phases of nonlinear systems, in the desire to understand how to generalize the well-developed BBC and topological characterization from linear to nonlinear systems. There are two types of nonlinear systems. The first type is the widely studied nonlinear-eigenvector systems. It emerges naturally in topological photonics in the presence of the Kerr effect \cite{PhysRevLett.117.143901,10.1063/1.5142397,Sone2024,Szameit2024,PhysRevLett.134.093801}, topolectric circuits by adding nonlinear electronic components \cite{Hadad2018,doi:10.1073/pnas.2106411118,PhysRevResearch.5.L012041}, and mechanical metamaterials of masses connected by nonlinear springs \cite{PhysRevB.100.014302}. The impact of nonlinearity on boundary states  gives rise to unique topological phenomena intertwined with the soliton \cite{PhysRevLett.100.013905,Zhang2020,PhysRevX.11.041057,PhysRevLett.128.093901,PhysRevB.107.184313,Choi2024,coen2024nonlinear} and synchronization \cite{PhysRevResearch.4.023211,Moille2025,https://doi.org/10.1002/advs.202408460}, and a potential realization of high-performance topological lasers \cite{Leefmans2024}. The BBC of the topological phases in this type of nonlinear systems has been constructed by directly defining topological invariants on the nonlinear eigenvectors \cite{Sone2024,doi:10.1126/science.aba8725}. The second type is the less explored nonlinear-eigenvalue systems \cite{Guttel_Tisseur_2017}. It occurs in photonic systems in a medium with frequency-dependent permittivity \cite{PhysRevB.50.16835,PhysRevA.78.033834,8937095,8358010,PhysRevE.106.035304,LI2022108835,photonics10050523,PhysRevA.109.043518}, in elastic metamaterials \cite{PhysRevB.110.174305}, and quantum open and many-body systems \cite{PhysRevA.94.022105,PhysRevB.111.045137}. It has been found that the topological phases of this type of systems cannot be uniquely determined by the Hamiltonian alone. Furthermore, their eigenvalues may become complex even in the Hermitian case. They make the exploration of the BBC and topological characterization of their topological phases troublesome. Only recently, a method was proposed to reveal the BBC of the real-band topological phases of a nonlinear-eigenvalue system by mapping the system to an auxiliary linear system \cite{PhysRevLett.132.126601}. However, an open question remains regarding nonlinear-eigenvalue systems with topological phases, particularly photonic systems, which typically possess non-Hermiticity caused by optical gain, loss, and nonreciprocity effects \cite{PhysRevLett.121.213902,Leefmans2024}. This non-Hermiticity can lead to a breakdown of the BBC \cite{PhysRevLett.116.133903,PhysRevLett.120.146402,PhysRevLett.121.086803,PhysRevLett.123.066404,PhysRevLett.124.086801,PhysRevLett.124.056802}. Thus, it is highly desirable to solve the problem of how to characterize the non-Hermitian topological phases in nonlinear-eigenvalue systems.

Here, we investigate the topological phase transition of a class of non-Hermitian nonlinear-eigenvalue systems with second- and third-order nonlinearities. By using the auxiliary-system method, a complete BBC and topological characterization of the non-Hermitian topological phases in such nonlinear systems is established. A non-Bloch band theory is developed to describe the non-Hermitian topological phases with nonlinear eigenvalues by introducing the generalized Brillouin zone on the auxiliary system. Importantly, we find that an exotic topological phase with coexisting real- and complex-band boundary states emerges in our non-Hermitian nonlinear-eigenvalue system. Our result serves as a useful tool for investigating the rich topological phases in non-Hermitian nonlinear-eigenvalue systems.

\begin{figure}[tpp]
\begin{center}
\includegraphics[width=\columnwidth]{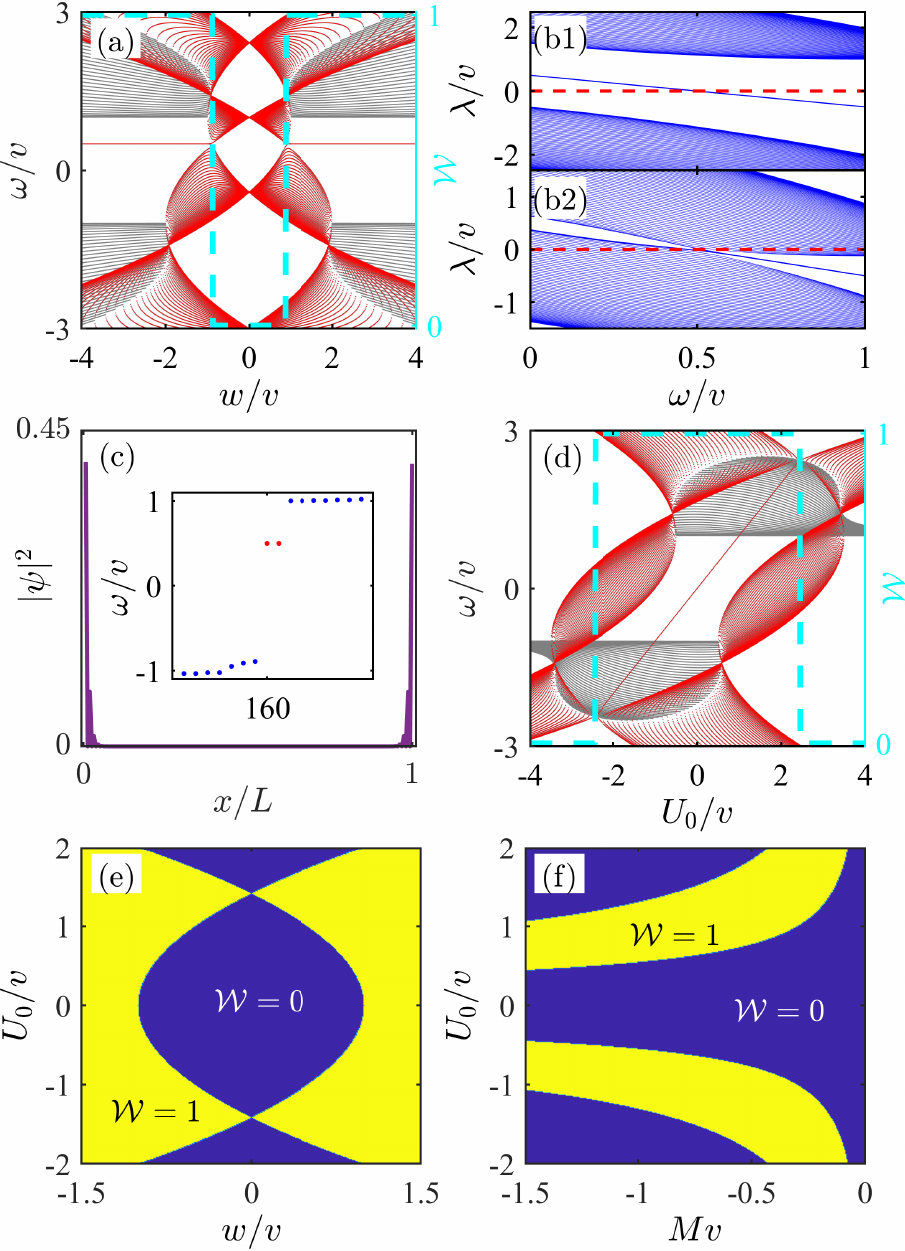}
\end{center}
\caption{(a) Real-$\omega$ bands (red lines), real part of complex-$\omega$ bands (gray lines), and winding number $\mathcal{W}$ (cyan dashed lines) in different $w$. Band structure of $\lambda$ when (b1) $w=2v$ and (b2) $0.875v$. (c) Probability distribution of the edge states, with the inset being the eigenvalues. (d) Real-$\omega$ bands (red lines), real part of complex-$\omega$ bands (gray lines), and $\mathcal{W}$ in different $U_0$ when $w=2v$. Phase diagrams in the (e) $U_0$-$w$ and (f) $U_0$-$M$ spaces when $w=0.75v$. We use $U_0=0.5v$ in (a)--(c) and $M=-0.5v^{-1}$ and $L=80$ for all.} \label{F1}
\end{figure}

\section{Topological phases in Hermitian nonlinear systems}
First, we consider a system satisfying a Hermitian nonlinear-eigenvalue equation
\begin{equation}
H_0\,|\psi \rangle=\omega\mathcal{S}(\omega)|\psi \rangle, \label{NH}
\end{equation}
where $H_0$ is a Hermitian Hamiltonian, $\mathcal{S}(\omega)$ is called the overlap matrix, and $\omega$ is the nonlinear eigenvalue. $\mathcal{S}(\omega)$ makes the system nonlinear. Equation \eqref{NH} can describe the light propagation in photonic crystals, where the frequency dependence of permittivity results in the presence of $\mathcal{S}(\omega)$ \cite{PhysRevB.50.16835,PhysRevA.78.033834,8937095,8358010,PhysRevE.106.035304,LI2022108835,photonics10050523,PhysRevA.109.043518}. The nonlinearity renders some of the eigenvalues $\omega$ to be complex even when $H_0$ is Hermitian \cite{52wp-42b4}. The system hosts the topological phases defined in the bands formed by all real $\omega$ \cite{PhysRevLett.132.126601,PhysRevB.109.134201}. The topological phases do not obey the well-established BBC in linear systems. In order to establish their BBC, we introduce an auxiliary system whose eigenequation reads $P(\omega)|\phi\rangle=\lambda|\phi\rangle$ \cite{PhysRevLett.132.126601,PhysRevA.78.033834,PhysRevA.111.042201}, where $P(\omega)=H_0-\omega\mathcal{S}(\omega)$ is called the pencil matrix \cite{Ikramov1993} and $\omega$ is treated as a free parameter. Then, running $\omega$ in a physically permitted regime for given system-parameter values and solving the eigen solution of $P(\omega)$, we obtain the $\lambda$ spectrum of the auxiliary system. The $\lambda$ spectrum has no physical meaning except for $\lambda=0$, where Eq. \eqref{NH} is recovered. Therefore, we succeed in converting the nonlinear-eigenvalue problem into a linear one. Such a conversion is exact. Inheriting the complete $\omega$ spectrum, the zero line of the $\lambda$ spectrum of the auxiliary system shares the same topological feature as the original system. When the $\lambda$ bands are topological, they possess gapless edge states under the open-boundary condition (OBC). These states cross $\lambda=0$ so that they also emerge in the original system \cite{PhysRevLett.132.126601}. Thus, we can use the topological invariant defined in the auxiliary system to characterize the edge states and reveal the BBC of the original system.

We investigate a one-dimensional second-order nonlinear Su-Schrieffer-Heeger (SSH) model with $H_0=\sum_{n=1}^L\left[U_0c^\dagger_{n}c_{n}+\left(vc^\dagger_{n,A}c_{n,B}+\text{H.c.}\right)\right]+\sum_{n=1}^{L-1}\left(wc^\dagger_{n+1,A}c_{n,B}+\text{H.c.}\right)$ \cite{asboth2016short} and $\mathcal{S}(\omega)=s_0+s_1\omega$, with $s_0=1$ and $s_1=-M\sum_{n=1}^L\left(c_{n,A}^\dag c_{n,B}+\text{H.c.}\right)$ \cite{52wp-42b4}. Here, $U_0$ is the on-site potential, $v$ and $M$ are the linear and nonlinear intracell hopping rates, and $w$ is the intercell hopping rate. Equation \eqref{NH} is rewritten in a linear form \cite{Jarlebring2012}
\begin{equation}
\left( \begin{array}{cc}0 & I  \\ s_{1}^{-1}H_0 & -s_{1}^{-1}s_0 \end{array}\right)\left( \begin{array}{c}|\psi\rangle \\
\omega |\psi\rangle \end{array}\right)=\omega\left( \begin{array}{c}|\psi\rangle   \\ \omega |\psi\rangle \end{array}
\right).\label{linar}
\end{equation}
By numerically solving Eq. \eqref{linar}, we obtain the four-band structure of $\omega$ under the OBC [see Fig. \ref{F1}(a)]. It shows that the real part of the complex-$\omega$ bands does not contribute to a topological phase transition and the band touching of the real-$\omega$ bands results in the presence of the gaped edge states with an eigenvalue $\omega=U_0$. The band structure is recovered by the eigenvalue spectrum of $\lambda$ of the auxiliary system under the OBC, where the gapless edge states cross $\lambda=0$ just at $\omega=U_0$ [see Fig. \ref{F1}(b1)]. Under the periodic-boundary condition (PBC), $P(\omega)$ is converted into the momentum-space form $P(k,\omega)=\sum_{j=0,\pm}p_j(k,\omega)\sigma_j$, where $\sigma_0$ is the identity matrix, $\sigma_\pm=(\sigma_x\pm i\sigma_y)/2$, $p_0(k,\omega)=U_0-\omega$, and $p_\pm(k,\omega)=v+M\omega^2+we^{\mp ik}$. It is easily obtained from $P(k,\omega)$ that, at the high-symmetry points $k=0$ and $\pi$, the bulk bands touch at $\omega=U_0$ when $MU^2_0+v=\pm w$, which gives the critical points of the topological phase transition. They match the ones obtained under the OBC, where the bands across $\lambda=0$ touch together just at $\omega=U_0$ [see Fig. \ref{F1}(b2)]. Such a matching firmly establishes the BBC of the original system. Defined in $H_0$ and $\mathcal{S}(\omega)$, the symmetries of the original system are fully preserved in $P(k,U_0)$. It satisfies time-reversal symmetry $\mathcal{T}P(k,U_0)\mathcal{T}^{-1}=P(-k,U_0)$, chiral symmetry $\mathcal{C} P(k,U_0)\mathcal{C}^{-1}=-P(k,U_0)$, and particle-hole symmetry $\Gamma P(k,U_0)\Gamma^{-1}=-P(-k,U_0)$, with $\mathcal{T}$ being the complex conjugation, $\mathcal{C}=\sigma_z$, and $\Gamma=\sigma_z\mathcal{K}$. Thus, the edge states exhibit the twofold Kramers degeneracy [see Fig. \ref{F1}(c)] and the topology is well described by the winding number
\begin{equation}
    \mathcal{W}={1\over 2\pi i}\int_{-\pi}^\pi {\rm d}k {{\rm d}\over {\rm d}k}\ln p_-(k,U_0).\label{wdn}
\end{equation}
It is easy to find $\mathcal{W}=1$ when $|MU_0^2+v|< |w|$.

We see from Fig. \ref{F1}(a) that $\mathcal{W}$ defined in the auxiliary system correctly characterizes the topological phases of the original system. Due to the nonlinearity, the on-site potential $U_0$ also plays a dominant role in the topological phase transition, which is different from linear systems. Figure \ref{F1}(d) displays the band structure of $\omega$ under the OBC in different $U_0$ when $w=2v$. Being similar to Fig. \ref{F1}(a), the complex-$\omega$ bands do not contribute any topological phase transition. The twofold degenerate edge states with $\omega=U_0$ are present in the real-$\omega$ bands when $U_0^2<-(w+v)/M$. They are well characterized by the winding number $\mathcal{W}$ defined in $P(k,U_0)$. To gain a global picture, we plot in Figs. \ref{F1}(e) and \ref{F1}(f) the phase diagrams in the $U_0$-$w$ and $U_0$-$M$ spaces. Figure \ref{F1}(f) demonstrates the nonlinearity term $M$ can drive the topologically trivial linear system with $M=0$ to a topological phase. It indicates that the nonlinearity supplies us an extra dimension to engineer the topological phases. Thus, we have established a complete topological characterization and the BBC for the Hermitian topological phases in the nonlinear-eigenvalue system, where the system topology is contributed by real bands. Next, we investigate the cases where non-Hermiticity is present in the Hamiltonian matrix and the overlap matrix, respectively.

\section{Real-band topological phases in non-Hermitian nonlinear systems}
Consider that the non-Hermitian terms induced by the non-reciprocal hoppings are present as
\begin{equation}
H=H_{0}+\delta \sum_{n=1}^L\left(c^\dagger_{n,A}c_{n,B}-c^\dagger_{n,B}c_{n,A}\right), \label{NR}
\end{equation}
where $\delta$ is a non-reciprocal parameter between the two directions of the intracell hopping rates \cite{PhysRevLett.123.066404}. $\mathcal{S}(\omega)$ takes the same form as the one in the preceding section. The pencil matrix of the auxiliary system under the PBC becomes $P(k,\omega)=\sum_{j=0,\pm}p_j(k,\omega)\sigma_j$, where $p_0(k,\omega)=U_0-\omega$ and $p_\pm(k,\omega)=M\omega^2+v+we^{\mp ik}\pm \delta$. It is easy to see that, at the two high-symmetry points $k=0$ and $\pi$, the bands of $P(k,\omega)$ touch at $\omega=U_0$ when $M U_0^2+v=w\pm \delta$ or $-w\pm \delta$. It is completely different from the band-touching condition $MU_0^2+v=\pm\sqrt{w^2+\delta^2}$ obtained under the OBC. This reflects the non-Hermiticity caused breakdown of BBC. In order to establish the BBC, we replace $e^{ik}$ by $\beta=re^{ik}$ with $r=\left[\left|\left(v+MU_0^2-\delta\right)/\left(v+MU_0^2+\delta\right)\right|\right]^{1/2}$, which defines a generalized Brillouin zone (BZ) \cite{PhysRevLett.121.086803}. Then $P(k,U_0)$ is converted into $\bar{P}(k,U_0)=\bar{p}_+(k,U_0)\sigma_++\bar{p}_-(k,U_0)\sigma_-$, with $\bar{p}_\pm(k,U_0)=MU_0^2+v+w
\beta^{\mp1}\pm\delta$. It is remarkable to find that its band-touching condition successfully recovers the one under the OBC. In the similar manner as Eq. \eqref{wdn}, we define the winding number of $\bar{P}(k,U_0)$ via the phase changes of $\bar{p}_\pm(k,U_0)$ when $k$ runs over the generalized BZ \cite{PhysRevLett.123.066404,PhysRevB.111.064310,PhysRevB.97.045106}
\begin{equation}
    \mathcal{W}={1\over 4\pi i}\int_{-\pi}^\pi {\rm d}k {{\rm d}\over {\rm d}k}\ln {\bar{p}_-(k,U_0)\over \bar{p}_+(k,U_0)}.\label{wdnhm}
\end{equation}
We find that $\mathcal{W} = 1$ and two edge states are formed in both the auxiliary and original systems when $\left|MU_0^2+v\right|<\sqrt{w^2+\delta^2}$.

\begin{figure}[tpp]
\includegraphics[width=\columnwidth]{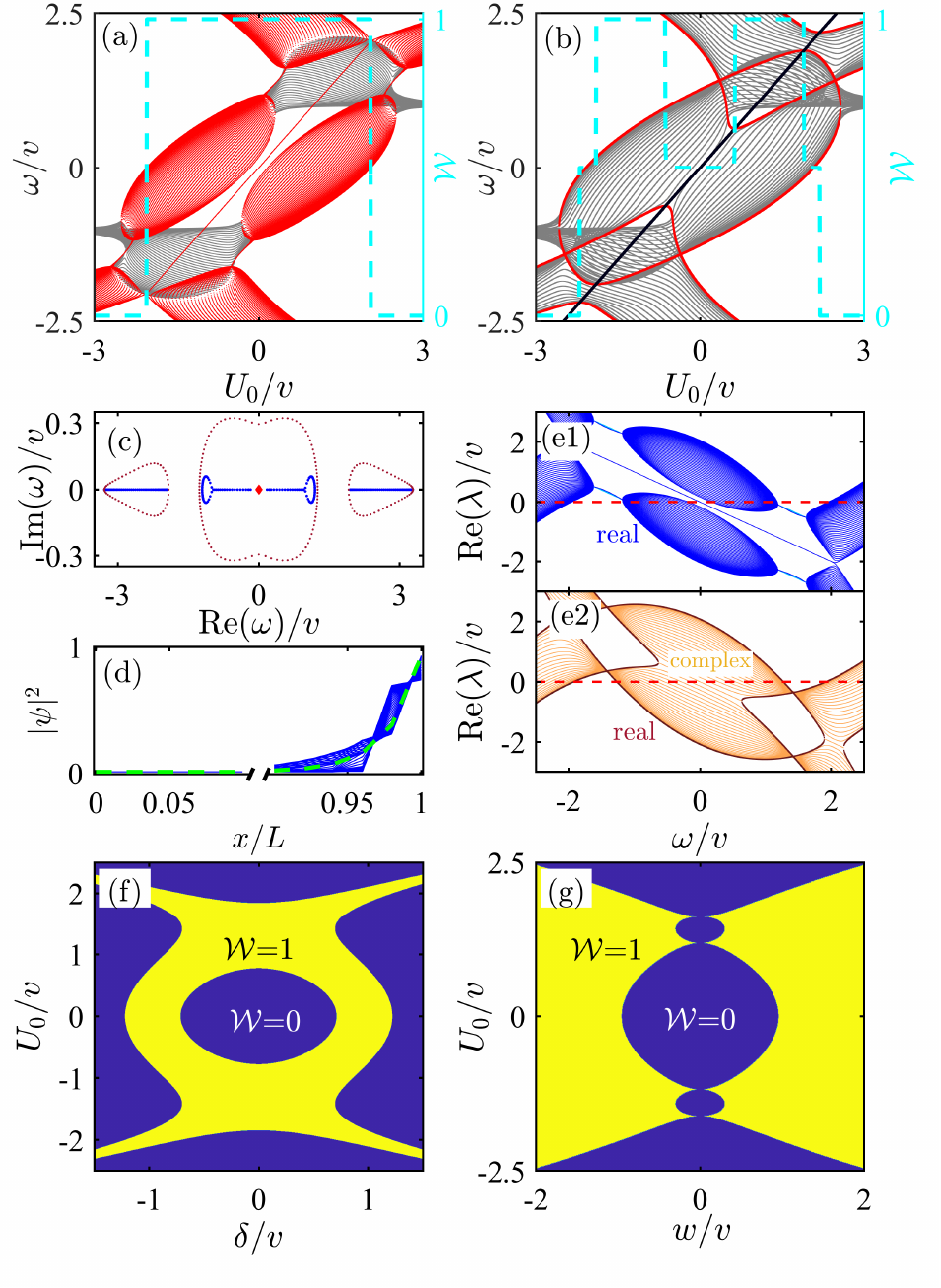}
\caption{Real-$\omega$ bands (red lines), real part of complex-$\omega$ bands (gray lines) under the (a) OBC and (b) PBC. Winding numbers $\mathcal{W}$ (cyan dashed lines) defined in the (a) generalized BZ and (b) BZ. The black solid line in (b) shows the Fermi level $\omega=U_0$. (c) Eigenvalues of $\omega$. (d) Distributions of the edge state (the green dashed line) and the bulk state (the blue line). Band structures of $\text{Re}(\lambda)$ under (e1) the OBC (blue) and (e2) PBC (brown) when $U_0=0$. Phase diagrams in the (f) $U_0$-$\delta$ and (g) $U_0$-$w$ spaces. We use $w=1.1v$ in (a)-(f) and $0.75v$ in (g) and $\delta=0.3v$, $M=-0.5v^{-1}$, and $L=50$ for all.} \label{F2}
\end{figure}

Figure \ref{F2}(a) shows the band structures of $\omega$ under the OBC in different $U_0$ obtained by numerically solving Eq. \eqref{linar}. Again, the topological phase transition is not contributed by the real part of the complex-$\omega$ bands, but by the real-$\omega$ bands. Accompanying the touching of the real-$\omega$ bands, the twofold degenerate edge states with $\omega=U_0$ protected by the time-reversal symmetry are present when $U_0^2<-\left[\sqrt{w^2+\delta^2}+v\right]/M$. This band structure is dramatically different from the one under the PBC [see Fig. \ref{F2}(b)], where the real-$\omega$ bands touch the Fermi level $\omega=U_0$ at $U_0^2=(w- \delta-v)/M$ or $(-w\pm \delta-v)/M$, while the bands of the real part of the complex $\omega$ are closed at almost all the positions where the edge states are present in Fig. \ref{F2}(a). The eigenvalues of $\omega$ when $U_0=0$ in Fig. \ref{F2}(c) confirm that the bands in the presence of the edge state under the OBC are closed under the PBC. The mismatching of the bulk band structures under these two boundary conditions clearly demonstrates the breakdown of the BBC caused by the non-Hermiticity. The distribution of the eigenstates in Fig.~\ref{F2}(d) shows that both the edge state and the bulk states accumulate toward the boundary, which shows the non-Hermitian skin effect. The two real-$\lambda$ band structures can be recovered by $\text{Re}(\lambda)$ of the auxiliary system under the corresponding boundary conditions in Fig. \ref{F2}(e1) for OBC and Fig. \ref{F2}(e2) for PBC. The gapless edge states in the $\text{Re}(\lambda)$ bands under the OBC cross $\lambda = 0$ precisely at $\omega = U_0$, which is consistent with Fig. \ref{F2}(a). However, the $\text{Re}(\lambda)$ bands under the PBC across $\lambda=0$ do not exhibit a gap. It verifies the breakdown of the BBC. The winding number of $P(k,U_0)$ calculated in the conventional BZ is given in Fig. \ref{F2}(b). Although qualitatively capturing the touching points of the real $\omega$ and the Fermi level $U_0$ under the PBC, the winding numbers defined in the conventional BZ nonphysically take half integers. Importantly, it cannot characterize the edge states under the OBC. However, the one defined in $\bar{p}_\pm(k,U_0)$ of the generalized BZ correctly counts the pair number of the edge states [see Fig. \ref{F2}(a)]. We thus establish a non-Bloch BBC for our nonlinear system. Figures \ref{F2}(f) and \ref{F2}(g) show the phase diagrams characterized by $\mathcal{W}$ in the $U_0$-$\delta$ and $U_0$-$w$ spaces. They give a global picture of the topological phases in the non-Hermitian nonlinear system.

We have discovered in this case that non-Hermiticity leads to the breakdown of the BBC in nonlinear-eigenvalue systems. Due to the topological inheritance of the original system by the auxiliary system under OBC, we can introduce a generalized BZ in the auxiliary system to restore the BBC. However, this still falls within the realm of real-band topology. Next, we will examine the scenario where non-Hermiticity is present in the overlap matrix such that the topological phases coexist in both the real and complex bands.

\begin{figure}[tpp]
\begin{center}
\includegraphics[width=\columnwidth]{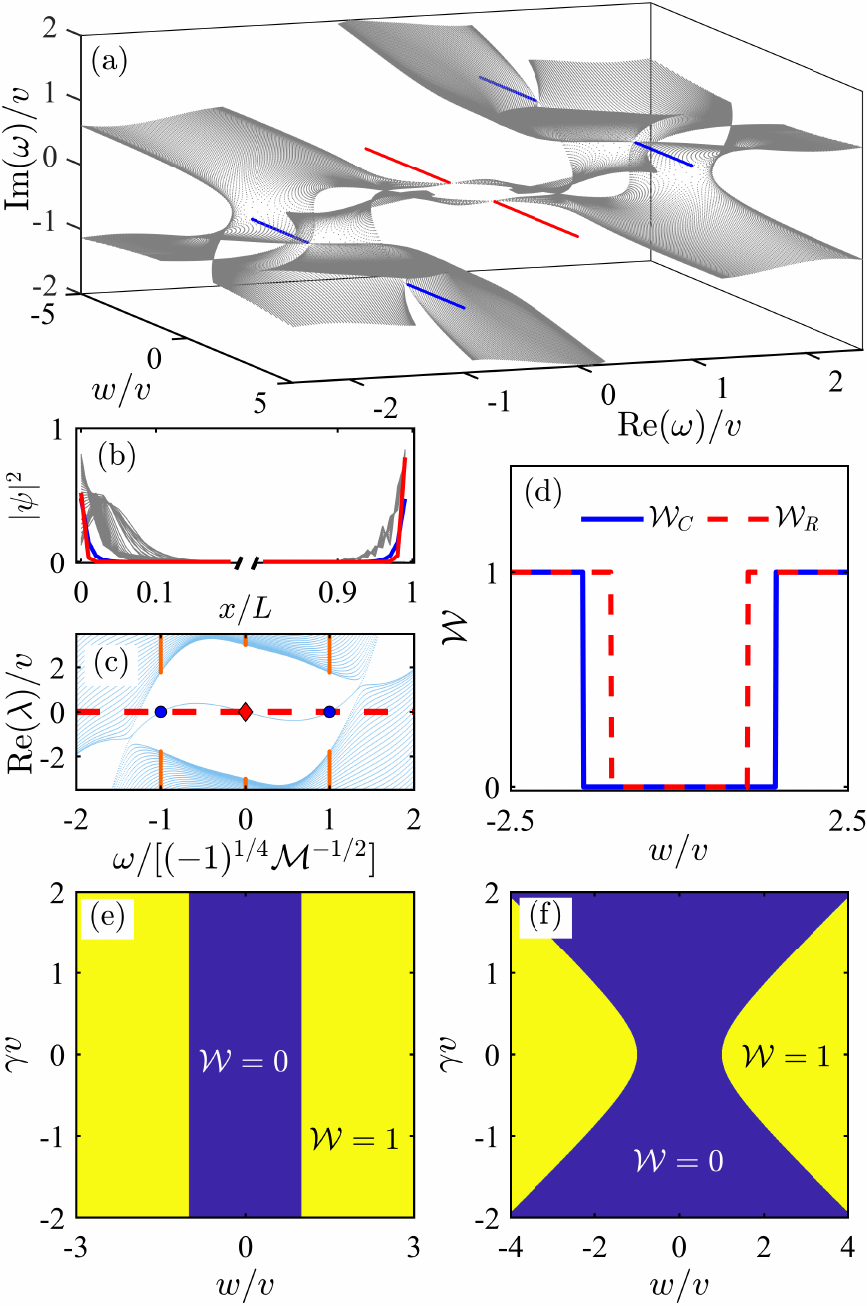}
\end{center}
\caption{(a) Band structure of $\omega$ in different $w$. The real and complex edge-mode eigenvalues are marked by red and blue lines, respectively. (b) Distribution of the real- (the red line) and complex-band (blue line) edge states and the bulk state (gray line). (c) Band structures of the real part of complex $\lambda$ (blue dots) and real $\lambda$ (orange dots) when $w=4v$. (d) Winding numbers in different $w$. Phase diagrams for the real (e) and complex (f) edge modes. We use $\mathcal{M}=-0.5 v^{-2}$, $\gamma=v^{-1}$, and $L=100$.}\label{F3}
\end{figure}

\section{Complex-band topological phases in non-Hermitian nonlinear systems}
We study a third-order nonlinear non-Hermitian SSH model. It possesses the same $H_0$ as above but with $U_0=0$ and a non-Hermitian $\mathcal{S}(\omega)=s_0+\tilde{s}_1\omega+\tilde{s}_2\omega^2$, where $s_0=1$, $\tilde{s}_1=-\gamma\sum_{n=1}^L\left(c_{n,A}^\dag c_{n,B}-c_{n,B}^\dag c_{n,A}\right)$ and $\tilde{s}_2=-i\mathcal{M}$ \cite{52wp-42b4,PhysRevLett.121.086803}. Here, $\gamma$ is a non-reciprocal parameter between the two
directions of the nonlinear intracell hopping rates and $\mathcal{M}$ is the nonlinear loss-gain parameter. Equation \eqref{NH} is rewritten as \cite{Jarlebring2012}
\begin{equation}
\left( \!\! \begin{array}{ccc}0 & I & 0 \\0 & 0 & I \\ \tilde{s}_{2}^{-1}H_0 & -\tilde{s}_{2}^{-1}s_0 & -\tilde{s}_{2}^{-1}\tilde{s}_1 \end{array}\! \! \right) \!\!  \left(\!\begin{array}{c}|\psi\rangle \\
\omega |\psi\rangle \\\omega^2 |\psi\rangle \end{array} \! \right)\!\! = \omega \! \left(\! \begin{array}{c}|\psi\rangle \\ \omega |\psi\rangle \\ \omega^2 |\psi\rangle \end{array}
\!\right). \label{linar3}
\end{equation}
The presence of $\tilde{s}_2$ turns the system into a six-band model. The pencil matrix is derived to be $P(k,\omega)=\sum_{j=0,\pm}p_j(k,\omega)\sigma_j$, where $p_0(k,\omega)=i\mathcal{M}\omega^3-\omega$ and $p_\pm(k,\omega)=v\pm\gamma \omega^2+we^{\mp ik}$. $P(k,\omega)$ satisfies the chiral symmetry $\mathcal{C} P(k, \omega) \mathcal{C}^{-1} = -P(k, -\omega)$. Its bands touch under the condition $i \mathcal{M}\omega^3 - \omega = 0$, which gives $\omega=0$ and $\pm(-1)^{3/4}\mathcal{M}^{-1/2}$. Thus, the system hosts both the real edge modes at $\omega=0$ and the complex edge modes at $\omega=\pm(-1)^{3/4}\mathcal{M}^{-1/2}$. For the former, the pencil matrix reduces to $P_{\rm R}(k,0)=\left(v+we^{-ik}\right)\sigma_++\text{H.c.}$, which is just a linear SSH model. At the high-symmetry points $k=0$ and $\pi$, the real bands of $P_{\rm R}(k,0)$ touch at $|v|=|w|$. The pencil matrix for the complex edge modes becomes $P_{\rm C}\left[k,\pm(-1)^{3/4}\mathcal{M}^{-1/2}\right]=\left(v-{i\gamma\over\mathcal{M}}+we^{-ik}\right)\sigma_++\text{H.c.}$, whose bands touch at $|k|=|\arccos[\pm v/\sqrt{(\gamma/\mathcal{M})^2+v^2}]|$ when $w^2=(\gamma/\mathcal{M})^2+v^2$. It is interesting to find that, although the system is non-Hermitian and the edge modes possess complex eigenvalues, the two pencil matrices $P_\text{R}(k,0)$ and $P_\text{C}\left[k,\pm(-1)^{3/4}\mathcal{M}^{-1/2}\right]$ of the auxiliary system are Hermitian. Therefore, the original and the auxiliary systems obey the BBC. The winding numbers for the real and complex bands can be defined in $P_{\rm R}(k,0)$ and $P_{\rm C}\left[k,\pm(-1)^{3/4}\mathcal{M}^{-1/2}\right]$ in a manner similar to that of the Hermitian case.  Without playing any role in the real-band topological phase, the nonreciprocity $\gamma$ and the gain/loss $\mathcal{M}$ together contribute to the complex-band topological phases. $\mathcal{M}$ gives the eigenvalues $\omega=\pm (-1)^{3/4}\mathcal{M}^{-1/2}$ of the topological edge states, while $\gamma$ impacts the critical point of the topological phase transition and the probability distribution of the edge states. First, the Hermiticity of $P_\text{C}\left[k,\pm(-1)^{3/4}\mathcal{M}^{-1/2}\right]$ means that the complex-band edge modes have the same BBC as the ones in the Hermitian systems. Second, well protected by the band gaps and chiral symmetry, these modes are robust to the fluctuation of the system parameters. Third, possessing complex eigenvalues, they are dynamically unstable.

By numerically solving Eq. \eqref{linar3}, we obtain the complex-$\omega$ bands under the OBC in different $w$ [see Fig. \ref{F3}(a)]. It is remarkable to find that the gapped edge states have not only a real eigenvalue $\omega=0$ (red lines) but also two complex eigenvalues $\omega=\pm(-1)^{3/4}\mathcal{M}^{-1/2}$ (blue lines). The former ones are present when $|w|>|v|$. The latter ones are present when $w^2>(\gamma/\mathcal{M})^2+v^2$. The distribution of the bulk state with complex eigenvalue in Fig.~\ref{F3}(b) shows the bulk states also reside in the edge, indicating the non-Hermitian skin effect. Figure \ref{F3}(c) shows the $\lambda$ band of the auxiliary system under the OBC. It clearly indicates that a gapless complex edge mode intersects with $\lambda = 0$ just at $\omega=0$ and $\pm (-1)^{3/4}\mathcal{M}^{-1/2}$, corresponding exactly to the two types of edge modes in Fig. \ref{F3}(a). The orange points in the bulk bands reveal that all the values of $\lambda$ are real, whose positions exactly match $\omega=0$ and $\pm (-1)^{3/4}\mathcal{M}^{-1/2}$. Thus, the auxiliary system at these three positions is Hermitian and we can calculate the topological numbers defined in $P_{\rm R}(k,0)$ and $P_{\rm C}(k,\pm(-1)^{3/4}\mathcal{M}^{-1/2})$ to characterize the real and complex edge modes, respectively. The obtained winding numbers correctly count the pair number of the two types of edge states [see Fig. \ref{F3}(d)]. The phase diagrams for the real and complex edge bands in the $\gamma$-$w$ space in Figs. \ref{F3}(e) and \ref{F3}(f) give a global picture of the non-Hermitian topological phases of this third-order nonlinear-eigenvalue system.

\section{Discussion and conclusion}
Although progress on topological phases in nonlinear-eigenvalue systems is rare, various systems, especially metamaterials, show great potential for realizing our predictions. Nonlinear-eigenvalue problems arise in various physical systems. In the study of higher-order nonlinear responses of graphene \cite{photonics10050523,SONG2020109871,photonics10121297}, the wave propagation problem is reduced to an eigenvalue problem. In analyzing waveguide filter structures with frequency dependent material properties or specific boundary conditions, the frequency-dependence of their permittivity or permeability make the analysis of their resonant characteristics to be a nonlinear-eigenvalue problem \cite{PhysRevB.50.16835,PhysRevA.78.033834,8937095,8358010,PhysRevE.106.035304,LI2022108835,photonics10050523,PhysRevA.109.043518}. Furthermore, waveguide arrays can also be used to simulate topological insulators \cite{ElHassan2019,Yang2023}, including non-Hermitian topological insulators \cite{PhysRevResearch.6.023140,PhysRevLett.133.073803,PhysRevLett.123.165701}. In the field of structural acoustics, the dynamic equation takes the form of a quadratic nonlinear-eigenvalue problem \cite{Chaigne2014}. Acoustic systems can be used similarly to model topological insulators \cite{PhysRevApplied.18.034066,PhysRevLett.128.224301,PhysRevX.11.011016,PhysRevLett.127.214301,Xue2020}. Therefore, our system can be conceptually derived from the design principles of these practical systems.

We have investigated the non-Hermitian topological phases in the nonlinear-eigenvalue system. A complete topological characterization has been established by introducing an auxiliary system. It is remarkable to find that, in addition to the conventional non-Hermitian topological phases in the real bands, exotic complex-band non-Hermitian topological phases are present in the system. In the case that the nonreciprocal hoppings in the Hamiltonian matrix break the BBC of the real-band topological phases, we have succeeded in restoring the BBC via introducing the generalized BZ in the auxiliary system. When the nonreciprocal hoppings are present in the overlap matrix, coexisting complex- and real-band topological phases are discovered. Although processing a complex eigenvalue, the complex-band topological phases correspond to a Hermitian Hamiltonian of the auxiliary system. This enables us to apply the well established characterization in the Hermitian system to describe the complex-band topological phases. Enriching the family of nonlinear topological phases, our result lays a foundation for exploring novel nonlinear topological phases, especially in optical, acoustic, and elastic metamaterial systems.

\section{Acknowledgment}
The work is supported by the National Natural Science Foundation of China (Grants No. 124B2090, No. 12275109, No. 92576202, and No. 12247101), the Quantum Science and Technology-National Science and Technology Major Project (Grant No. 2023ZD0300904), the Fundamental Research Funds for the Central Universities (Grant No. lzujbky-2025-jdzx07), and the Natural Science Foundation of Gansu Province (Grants No. 22JR5RA389 and No. 25JRRA799).

\bibliography{BIB}
\end{document}